\documentclass[%
 reprint,
 amsmath,amssymb,
 aps, prl
]{revtex4-2}
\usepackage{graphicx}
\usepackage{textcomp}
\usepackage{amssymb}
\usepackage[colorlinks,linkcolor=blue,anchorcolor=blue,citecolor=blue,urlcolor=black]{hyperref}
\usepackage[utf8]{inputenc}
\usepackage{amsmath}
\usepackage{tipa}
\usepackage{comment}
\usepackage{tabularray}
\usepackage{mathtools}
\usepackage{braket}
\usepackage{listings}
\usepackage{xcolor}
\usepackage{soul}
\usepackage{etoolbox}
\usepackage{lineno}
\usepackage{ulem}
\usepackage{svg}
\usepackage{CJKutf8}
\usepackage{multirow}
\usepackage{tabularray}

\begin{document}

\title{Can Bose-Einstein condensates enhance radioactive decay?} 

\author{Hanzhen Lin (\begin{CJK*}{UTF8}{gbsn}林翰桢\end{CJK*})}
\affiliation{Department of Physics, Massachusetts Institute of Technology, Cambridge, MA 02139, USA}
\affiliation{Research Laboratory of Electronics, Massachusetts Institute of Technology, Cambridge, MA 02139, USA}
\affiliation{MIT-Harvard Center for Ultracold Atoms, Cambridge, MA, USA}

\author{Yu-Kun Lu}
\affiliation{Department of Physics, Massachusetts Institute of Technology, Cambridge, MA 02139, USA}
\affiliation{Research Laboratory of Electronics, Massachusetts Institute of Technology, Cambridge, MA 02139, USA}
\affiliation{MIT-Harvard Center for Ultracold Atoms, Cambridge, MA, USA}

\author{Wolfgang Ketterle}
\affiliation{Department of Physics, Massachusetts Institute of Technology, Cambridge, MA 02139, USA}
\affiliation{Research Laboratory of Electronics, Massachusetts Institute of Technology, Cambridge, MA 02139, USA}
\affiliation{MIT-Harvard Center for Ultracold Atoms, Cambridge, MA, USA}

\begin{abstract}
This paper lays out the principles of how Bose-Einstein condensates can modify radioactive decay. We highlight the challenges of many modes and short coherence times due to the $\approx$ MeV energies of the emitted radiation. Recent proposals for gamma ray and neutrino lasers claim that using a Bose-Einstein condensate as a source would solve these issues. We show that this is not the case, and the proposed experiments would have a gain of only $10^{-16}$ or smaller. We also analyze proposals for gamma ray lasers based on stimulated annihilation of positronium Bose-Einstein condensates.
\end{abstract}
\maketitle

There is growing overlap between nuclear physics and atomic and optical physics. Radioactive atoms and molecules are used or proposed for precision measurements and searches for dark matter~\cite{RevModPhys.90.025008}. Recently, a nuclear clock was realized~\cite{Nuclear_clock}, Hanbury Brown-Twiss correlations are identified in high-energy collision~\cite{HBT_2006}, and superradiance of 14.4 keV emission of $^{57}$Fe has been observed~\cite{Superradiance_Fe57}. Several papers suggest using Bose-Einstein condensates of radioactive atom or positronium to realize lasers or superradiant emission for beta rays, gamma rays or neutrinos~\cite{NAMIOT199656,HOPE199687,Gamma_Ray_Laser_Review,Avetissian2014,Avetissian2015,MARMUGI2018,neutrino_laser_jones_formaggio}. Here we examine several proposals over the last decade which claim that the coherence of Bose-Einstein condensates (BEC) provides a new mechanism for generating coherent beams of gamma rays or neutrinos~\cite{Avetissian2014,Avetissian2015,MARMUGI2018,neutrino_laser_jones_formaggio}. 
In an accompanying paper we prove that superradiant neutrino lasers are fundamentally impossible due to fermionic statistics~\cite{FermiImposible}. Here we focus on two general limitations of such proposals, the short coherence time due to nuclear recoil and the multi-mode geometry. As we show here, BECs can improve these limitations, but there are still 16 orders of magnitude missing to enhance radioactive decay using superradiance in contrast to recent claims~\cite{MARMUGI2018,neutrino_laser_jones_formaggio}. 

\begin{figure}[h]
    \centering
    \includegraphics[width= \linewidth]{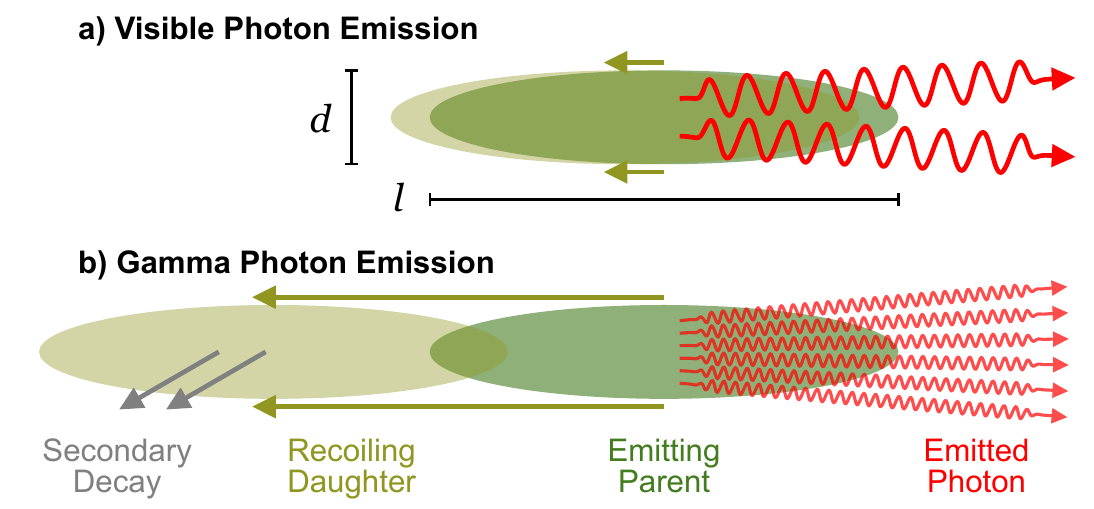}
    \caption{
   Superradiant ``laserlike'' emission of light has been observed from sodium BEC after laser excitation (a)\cite{SR_Rayleigh}. Recently, it has been suggested that radioactive BEC can be used to realize gamma ray~\cite{MARMUGI2018} and neutrino lasers~\cite{neutrino_laser_jones_formaggio} at MeV energies (b). Although superradiance of visible emission and gamma ray radiation are formally identical, the wavelength of MeV gamma rays is $10^6$ times shorter. As a result, many more photon modes (red) are populated, and the daughter atoms (olive) recoils faster and lose overlap with the parent atoms (blue) much sooner. In addition, the daughter atoms can be unstable and undergo secondary decays (gray), limiting the coherence time further.}
    \label{fig:interferometer}
\end{figure}

\textbf{Basic equations for gain and loss}
Since there have been widespread misconceptions on this topic for more than a decade, we first derive and discuss the principles of how atomic systems can modify radioactive decay before addressing specific proposals. Our discussion assumes a parent atom to decay into a recoiling daughter atom and a relativistic particle which escapes with the speed of light. For simplicity we refer to the emitted particle as a $\gamma$ ray photon, but our conclusions apply also to other particles.

In the End Matter we derive the following principle:

\noindent\textit{(1) Single-body loss cannot be modified by many-body physics and is always exponential.}\\
Here single-body loss is defined under the Born-Markov approximation as the sudden disappearance of single particles by a loss mechanism described by a  decay constant $\Gamma$  which characterizes the radioactive decay.  This proves that any form of constructive interference between different atoms is not possible.  Therefore, no special correlations of the parent atoms (BEC, Luttinger liquid) can modify radioactive decay~\footnote{Note that a recent paper~\cite{Luttinger_liquid_decay} claimed incorrectly that one-body decay is modified by a Luttinger liquid. The theoretical derivation uses an invalid long-wavelength approximation. It is not clear why the experimental results agree with the incorrect prediction. Our proof also rules out that there is bosonic enhancement of particle loss by optical pumping, as claimed in Ref.~\cite{vdStraten_PhysRevLett.116.173602}. }. The approximation for sudden loss  covers the situation where all decay products (e.g. recoil nuclei and emitted gamma rays or neutrinos) rapidly escape from the system.
The single-particle decay rate $\Gamma$ can include modification from the environment. 
The electron density affects the matrix element for electron capture processes~\cite{PhysRev.76.897,pressure_dependency_radioactive_decay_be7}. The atom's ionization state affects the rate of bound state beta decay \cite{PhysRevLett.77.5190}, and the density of states is modified in 
cavities and photonic bandgap materials, or via an index of refraction for photon emission~\cite{NIENHUIS1976181} as observed in thorium nuclear clocks~\cite{Nuclear_clock}.    
In these cases, our proof still applies with a renormalized single-particle loss rate.

In reality of course, single particles do not suddenly disappear, but have a finite coherence time with respect to the emitting system - this is non-Markovian dynamics. The full Hamiltonian for the particle decay includes the quantum fields for the photon and the daughter (ground state) atom~\cite{ketterle2001cargese}. For early times the evolution is fully coherent. Usually, there are finite coherence times for both the photon field (they escape the condensate volume typically in a few ps) and also for the daughter atom (they can escape due to recoil, decay to lower-lying states or dephase). The fastest decaying mode is adiabatically eliminated. If the atomic coherence decays faster than photon coherence, one retrieves the equation for optical gain, i.e. a laser medium. If the photon escapes quickly, one obtains the equations for superradiance~\cite{SR_Rayleigh,ketterle2001cargese}. In either case, one obtains a generic gain equation 
which says that the rate of spontaneous emission of photons into the relevant mode has to be faster than the loss rate for coherence. 

\noindent\textit{(2) The (renormalized) particle decay can only be enhanced by the presence of the decay products.}\\
The gain rate $G$ for the number of decay products $M$ within the condensate volume (photons for a laser medium, recoiling atoms or a spin-wave excitation in the emitting medium in the superradiant case) is the rate of emission into a preferred mode of the photon field, $G=N \Gamma \Omega$, where $N$ is the number of atoms, $\Gamma$ is the spontaneous rate (natural linewidth) of the transition emitting particle with wavelength $\lambda$, $\Omega \approx (\lambda/d)^2$ is the solid angle for coherent emission from a BEC of diameter $d$. We neglect additional angular terms due to e.g. a dipolar emission pattern. Expressing the loss rate of the coherence between initial and final atomic states as $L=1/\tau_{\mathrm{coh}}$, we obtain the rate equation $\dot M= G(M+1) – LM$ with the threshold condition for exponential amplification as $N \Gamma (\lambda/d)^2 \tau_{\mathrm{coh}} > 1$~\cite{SR_Rayleigh}.

This equation generalizes the simple Dicke model to extended samples. The collective decay rate $N \Gamma$ is reduced by the solid angle of coherent emission~\cite{GrossHaroche_superradiance,quantum_theory_superradiance_french}. For a Fresnel number $F=\pi d^2/4l \lambda \approx 1$~\cite{quantum_theory_superradiance_french} superradiance occurs essentially in a single mode.  For the case considered here (short wavelength) $ F \gg 1$, there are $\approx F^2$ end-fire modes with the same gain. Each of these end-fire modes have little diffraction and maintain coherence along the entire length of the sample~\cite{GrossHaroche_superradiance,quantum_theory_superradiance_french}, and one obtains the same threshold condition. 
For small $F$, the effective gain length $l$ is reduced to $Fl$ since diffraction reduces the overlap of the endfire mode with the sample, and the threshold for superradiance is $1/F$ times higher.

Below threshold, the steady state solution (for an undepleted source) is $g\coloneqq M=G/(L-G)$ which enhances the emission into the preferred mode to $(1+g) G$ with a dimensionless gain $g$. At the threshold for superradiance, $g$ diverges.  Below  threshold, the (undepleted) system reaches a steady state  with a single pass gain of $1+g$ ~\cite{ketterle2001cargese}.

We now focus on the superradiant case analogous to Rayleigh superradiance in a BEC~\cite{SR_Rayleigh,ketterle2001cargese} since photons leave the condensate volume in 3~ps assumed to be faster than other decoherence processes. In general, $M$ is the number of daughter atoms which have maintained coherence with the emitting medium and can be regarded as the number of excitations of a bosonic mode describing phonons or spin-excitations. Formally, it represents the off-diagonal matrix element (coherence) of the density matrix for the emitting two-level system. 

\textbf{Particle emission creates entangled states in multiple modes.} We first discuss the solid angle factor which expresses that amplification can happen in many modes, dramatically lowering the gain. In idealized Dicke superradiance~\cite{DickePhysRev.93.99}, the extent of the atomic sample is smaller than the wavelength of emission, and all of the emission is single-mode and therefore fully coherent. In many cases, superradiance is realized in geometries $\gg \lambda$. This is necessarily the case for 
MeV radiation with $\lambda\sim \mathrm{pm}$, hence the gain is multi-mode. Usually, elongated samples are favorable for gain along the end-fire modes which have a coherent solid angle $\Omega \approx (\lambda/d)^2$ assuming a sample with diameter $d$ and length $l$~\cite{GrossHaroche_superradiance}. Side modes have a smaller gain due their smaller coherent solid angle of $\approx \lambda^2/dl$.

Several proposals argued that the coherence of the BEC implies that the whole emission is single-mode and omitted the $\lambda^2 /d^2$ factor, which overestimates the gain by $10^{12}$~\cite{neutrino_laser_jones_formaggio,MARMUGI2018}. 
Therefore we now derive this multi-mode factor from multiple perspectives.
Semi-classically, if the emitters in the sample constructively interfere in one direction, this constructive interference is restricted to a solid angle $\lambda^2/d^2$.
Alternatively, we can look at the transverse phase space volume $V_p$ with two spatial and two momentum dimensions for photons emitted from a sample of diameter $d$ with momentum $\hbar k$: $V_p \sim d^2 \hbar^2 k^2$. Semiclassically, the number of quantum states is $V_p/h^2 \sim d^2/\lambda^2$, which the photons occupy. This is consistent with the uncertainty principle in
the photon’s transverse momentum, as photons emitted
within an angle of $\lambda/d $ cannot be distinguished in the far field and should therefore be associated with the same mode.  We emphasize that the symmetry of a spherical BEC does not restrict radiation into a single spherical mode. It emits partial waves up to an angular momentum $j\sim d\times q \sim d/ \lambda$ ($q=\hbar k$ is the photon momentum), and with $2j+1$ sub-states for each $j$ the number of partial waves is again $\sim d^2 /\lambda^2$. 

A common belief is that starting with a BEC, unitary time evolution would keep the system in a single mode of photon emission with the atoms in a fully symmetric state. However, although the system evolves in a pure state,  it is an entangled state which is a coherent superposition of many product states. In each component the photon is emitted into a specific direction with an atomic excitation which recoil in the opposite direction. If we are interested only in the emitted radiation we trace out the atomic degrees of freedom and obtain an incoherent mixture of photons moving in different directions where each distinguishable direction is a mode. More formally, the number of modes is the rank of the Schmidt decomposition of the entangled pure state into the bi-partite system of photons and atomic states~\cite{ekert1995entangled}.

\noindent\textit{(3) Superradiance is only possible if the number of atoms exceeds the number of modes. In free space, this condition requires $N (\lambda/d)^2 >1$.}\\
The multi-mode geometry sets a stringent requirement for the number of atoms. During the coherence time, one or more photons must be emitted into one of the high gain modes (i.e. endfire modes): $N (\lambda/d)^2\Gamma \times \tau_{\mathrm{coh}}> 1$ or $N (\lambda/d)^2 = n \lambda^2 l >1/(\tau_{\mathrm{coh}} \Gamma)$ where $n=N/(d^2 l)$ is the atomic density. Since  $1/\Gamma$ is an upper bound for $\tau_{\mathrm{coh}}$ one obtains the necessary condiiton $n \lambda^2 l >1$.

Since $\approx \lambda^2$ is the unitarity limited cross section for absorption or stimulated emission (neglecting a small numerical factor), $n \lambda^2 l$ is often called the optical density (OD). For a typical BEC with 3 $\mu$m diameter and a wavelength of 1 pm, $\mathrm{OD}>1$ requires $N> 10^{13}$, or the emitter will be depleted before a second photon is emitted into any relevant mode. We emphasize that the criterion $\mathrm{OD}>1$ was derived from the gain competition of multiple modes and applies also to situations where the cross section is smaller than $\lambda^2$ or where the gain mechanism is different (see two-photon annihilation of positronium below).

 \textbf{The origin of the coherence time.}
 Several proposals assumed that the presence of a condensate will provide an (almost) infinite coherence time lasting seconds or even longer~\cite{MARMUGI2018,neutrino_laser_jones_formaggio}. In reality the coherence time for the proposed system is very short with the transit time of the recoiling atoms through the condensate being an upper limit.
 
In superradiance, the first photon (per mode) is emitted at the normal spontaneous rate. But then, as lucidly expressed by Dicke~\cite{Dicke1964} ``The memory of the previously emitted electromagnetic field is burned into the radiating system'' so that ``the probability of radiating a photon in a particular direction is given by the normal incoherent intensity multiplied by the total number of photons previously radiated in that direction plus 1.'' So what makes superradiance possible is the memory time of the radiating system. This requires the system to emit the second photon before the memory of the first photon emission is lost. This memory time is the decay time of the off-diagonal matrix element of the density matrix connecting the initial and final state of emission.

\noindent\textit{(4) Superradiant enhancement of radioactive decay has to occur faster than the coherence time. It is set by further decay, or in free space by Doppler broadening and transit time of the recoiling atom.}\\
The coherence time can be limited by multiple processes. In the case of a cascade decay, i.e. the daughter nucleus rapidly decays, it is this decay time (which is fs for neutrino emission, and 50 ps for $^{135m}\mathrm{Cs}$, see table \ref{table:experimental_parameter} and discussion of specific proposals). In the case of a recoiling daughter atom leaving the condensate, it is transit time (i.e. the length of the condensate divided by the recoil velocity, 0.3 $\mu$s for MeV photons). For a thermal (non-condensed cloud), it is the time for the recoiling atom to move by one thermal de-Broglie wavelength, which is identical to the inverse of the Doppler broadening $k \Delta v$ where $k$ is the momentum transfer and $\Delta v$ the velocity spread. Actually, by identifying the momentum uncertainty of the condensate as $\hbar/l$, the time for atoms to recoil out of the condensate can also be interpreted as the inverse Doppler width of the condensate~\cite{BraggPhysRevLett.82.4569}.

For any radioactive Bose-Einstein condensate, the transit time of the recoiling atom sets an upper limit for the coherence time. Combined with the solid angle factor, the superradiant gain $g$ is then inversely proportional to the photon energy cubed. Therefore, an experiment done with optical photons (in the eV range) is easily feasible, but it doesn't scale well towards higher energies. For MeV photons, the gain $g$ is down by 18 orders of magnitude. As we discuss below, in the two proposals, the coherence time is even shorter due to rapid decay of the daughter atoms.

 \textbf{Dicke memory has to be local.}

 \noindent\textit{(5) Stimulation of decay requires local overlap between the emitting atoms and the decay products.}\\ 
In the neutrino laser proposal~\cite{neutrino_laser_jones_formaggio}, the authors obtained a very long coherence time since they presented a model where even after the daughter atoms have undergone further decay, escaped and have hit a detector, they are still entangled with the remaining condensate atoms in a symmetric Dicke state and therefore superradiant decay will occur. We disagree. When energetic atoms interact with a wall or detector, they usually immediately (within ns in a gas, ps in solid state) lose their coherence and are distributed over many quantum states.

But let’s assume there exist detectors that can permanently trap the daughter atoms, and assume these daughters are bosons. Then, when the parent atoms decay, the daughter bosons in the detector could enhance the decay by bosonic stimulation (which is one special case of superradiance~\cite{BosStim2023,SR_Rayleigh}). However, since the decay operator is local, any stimulation would be absent if the daughter atoms have no overlap with the parent atoms.

This can be formally proven for a short-ranged transition operator. If we use a basis of localized states and perform the sum over all final states, we only get contributions from final states which have spatial overlap with the atoms in the initial state. Therefore, the decay rate for the total system is identical to an ensemble which is obtained by performing the partial trace over all states which don’t overlap with the condensate. After performing this partial trace, we can apply the rigorously proven principle (1) to the reduced density matrix stating that singe particle decay is immutable for any system consisting only of atoms in the parent state. The authors’ statement ``.... the correlations among trapped parent atoms within the decaying
condensate are sufficient to preserve SR enhancement''~\cite{neutrino_laser_jones_formaggio} is incorrect. Superradiance requires the local presence of atoms in the final state of the emission. Holes or correlations among the parent atoms do not modify radioactive decay, as we have rigorously proven. Particle-hole symmetry does not apply here, since holes do not carry memory in the Dicke sense. Therefore, the transit time of recoiling atoms is a rigorous upper limit to the coherence time. 

We now emphasize two other principles:\\
\noindent\textit{(6) The properties of a Bose-Einstein condensate are not relevant for emission of light or scattering of light. There is no condensate ``magic''.}\\
This immediately follows from our discussion above. The optical properties of a BEC are identical to a non-condensate medium with the same density and Doppler width (given by the velocity distribution)~\cite{KetterleMWA} (see also End Matters). The condensate suffers from the same limitations of emission into many modes and a short coherence time as a thermal cloud. There is no mechanism which automatically transfers the coherence of the condensate to the emitted photon and recoiling atoms. Otherwise, we would have already observed enhanced single-body decay in BECs of metastable Helium \cite{metastable_helium_bec_cohentannoudji,metastable_helium_bec_alainaspect}. In cold atom experiments, a condensate has usually higher gain than a thermal cloud (since it is smaller) and a longer coherence time (since its velocity spread is smaller), but these are quantitative and not qualitative differences. The situation is somewhat different for positronium annihilation lasers (see End Matters).

\noindent \textit{(7) Dicke superradiance works in principle for fermionic emitters, but not for fermionic radiation.}\\
The classical discussion of Dicke superradiance assumes localized emitters, hence the quantum statistics of the parent do not play any role. For a gas of fermionic emitters there is additional Doppler broadening and a shortening of the coherence time due to the Fermi momentum but super-radiance is still possible~\cite{KetterleMWA,Meystre_PhysRevLett.86.4199,Fermion_SR_PhysRevLett.106.210401}.
In a accompanying paper we show that superradiance is impossible when the the emitted particle is a fermion. Fermionic anti-commutators introduce minus signs and remove the constructive superposition of matrix elements~\cite{FermiImposible}.

\begin{table}
\centering
\begin{tblr}{
 width = \linewidth,
  colspec = {Q[300]Q[192]Q[450]},
  row{1} = {c},
  cell{1}{1} = {c=2}{0.492\linewidth},
  cell{2}{1} = {c=2}{0.492\linewidth},
  cell{3}{1} = {c=2}{0.492\linewidth},
  cell{4}{1} = {c=2}{0.492\linewidth},
  cell{5}{1} = {r=2}{},
  cell{7}{1} = {c=2}{0.492\linewidth},
  cell{8}{1} = {c=2}{0.492\linewidth},
  cell{9}{1} = {c=2}{0.492\linewidth},
  cell{10}{1} = {c=2}{0.492\linewidth},
  vlines,
  hline{1-5,7-11} = {-}{},
  hline{6} = {2-3}{},
}
Condensate number                         &                        & $10^6$(atom)                                                 \\
Condensate density n                      &                        & $10^{14}\mathrm{cm}^{-3}$ (atoms);                           \\
Length $l$                                &                        & 1 mm (atom)                                                  \\
Diameter $d$                              &                        & 3 $\mu$m                                                   \\
Decay half life $\ln(2)/\Gamma$           & $^{135m}\mathrm{Cs}$   & 53 min ($\gamma$, M4)  \newline then $\sim$50 ps ($\gamma$, E2)                     \\
                                          & $^{83}\mathrm{Rb}$     & 86 days ($e^-$ capture)  \newline then $\sim$1 fs (K shell em.)      \\
Wavelength $\lambda$~                     &                        & $1\sim2$pm (0.5-1 MeV)                                       \\
Single mode solid angle $\lambda^2/r^2$   &                        & $10^{-12}$                                                   \\
Photon transit time $l/c$                 &                        & 3 ps                                                         \\
Atom transit time $l/v_\mathrm{recoil}$          &                        & $0.3\sim0.5 \mu  s$                                          \\
\end{tblr}
\caption{Experimental parameters proposed to enhance nuclear decay of atoms in a BEC~\cite{MARMUGI2018,neutrino_laser_jones_formaggio}.}
\label{table:experimental_parameter}
\end{table}

\textbf{Specific Proposals}
We can now examine specific proposals for enhancement of nuclear decay using Bose-Einstein condensates. Suggested experimental parameters for the proposals to enhance gamma-ray emission from $^{135m}\mathrm{Cs}$~\cite{MARMUGI2018}, neutrino emission from $^{83}\mathrm{Rb}$ (neglecting its fermionic nature)~\cite{neutrino_laser_jones_formaggio} are summarized in Table \ref{table:experimental_parameter}. 
For photon energies around 1 MeV, the photon wavelength is on the order of pm. For a cloud of few $\mu$m diameter the coherent solid angle is on the order of $10^{-12}$. 

The authors of the gamma ray laser proposal predict that a condensate of $^{135m}\mathrm{Cs}$ will produce an enhancement of the decay rate by more than 3 orders of magnitude compared to the spontaneous lifetime of 53 min or 3000 s. The experimental realization of this proposal has received funding from two sources~\cite{Gamma_funding} and is a long-term goal for the trapping and cooling of cesium isotopes at the atom-trap facility at the University of Jyväskylä, Finland~\cite{PLATAN2024}. Using the equations above and the parameters in Table \ref{table:experimental_parameter}, we obtain a gain $g \approx 10^{-16}$ if we assume that coherence time is limited by the recoil of the daughter. However, the daughter atom decays further with a  estimated lifetime of 50 ps~\cite{MARMUGI2018}, which can reduce the gain to $g\approx N (\lambda/d)^2 \Gamma \times 50\, \mathrm{ps} \approx 10^{-20}$, unless there is two photon-superradiance due to coherence between the initial and final state of the cascaded gamma ray transition \cite{Two_photon}. 

The gamma-ray laser proposal~\cite{MARMUGI2018} is based on the incorrect assumption that the condensate has special properties: ``The collective de-excitation relies on the coherence of the condensate being transferred to $\gamma$ photons ...''; ``indistinguishable nuclei in the BEC imprint the coherence of the boson field in the photon field. In fact, this process could not happen in mere cold atomic samples.''\cite{MARMUGI2018} We disagree with these statements. The authors use the same coupled equations for atoms and light discussed above~\cite{ketterle2001cargese}, but assume infinite coherence time for both daughter atoms and gamma rays in the main text. In the supplement, they add the rapid secondary decay of the daughter, and after adiabatic elimination for this state obtain an equation which looks identical to the expression for a two-level optical amplifier where each atom has a gain cross section proportional to $\lambda^2$, and the gain is reduced by a factor $\Gamma/L$ due to the decay rate $L \gg \Gamma$ of the daughter atom.  However, the authors then assume that gamma rays and parent atoms have an infinite coherence time (due to the assumed special properties of the condensate), whereas in reality it is limited by the photon escape time to 3~ps. 

The neutrino laser proposal~\cite{neutrino_laser_jones_formaggio} derives equations with the conclusion that superradiance can reduce the decay lifetime for $^{83}\mathrm{Rb}$ atoms from 86 days ($10^7$ s) to 1 min, a gain of 5 orders of magnitude through collective decay of $N=10^6$ atoms. This proposal has a lot of similarities to the gamma laser, but with additional feature that the emitted neutrinos are fermions, making superradiance fundamentally impossible~\cite{FermiImposible}. The authors incorrectly claim that BEC would eliminate the restrictions of the multi-mode behavior, the recoil and subsequent decay. Neutrino emission occurs via electron capture and creates a $^{83}$Kr atom with a K shell vacancy. The 2p to 1s transition in hydrogen has a lifetime of 1.6 ns. Applying the $Z^{-4}$ scaling gives an estimated lifetime for the daughter atoms of around 1 fs in agreement with measurements~\cite{K_Shell}. This lifetime is shorter than the neutrino transit time of 3~ps and puts the system into a regime of single-pass amplification instead of superradiance~\cite{PhysRevX.6.011025}.  Applying our discussion on multi-mode gain competition and coherence time, we find the single-pass optical gain to be $g\approx N (\lambda/d)^2 \Gamma \times 3 \, \mathrm{ps} \approx 10^{-23}$.If the K shell emission can be regarded as coherent mixing of the initial and final states, then the analog of two-photon superradiance \cite{Two_photon} has a coherence time equal to the transit time of the recoiling atom with $g\approx 10^{-18}$. 

Both the gamma ray laser~\cite{MARMUGI2018} and neutrino laser proposals were inspired by prominent paper about positronium annihilation lasers~\cite{Avetissian2014} where it stated that a BEC of positronium would allow a coherent generation of annihilation radiation. It has been suggested for many years~\cite{LIANG1988419,MILLS2002} that a sufficiently dense elongated BEC of positronium would cause stimulated gamma ray emission~\cite{Dirac_1930,Varma1977,Stim_Annihil_1979} and therefore directional amplification of the annihilation radiation. The singlet (para) state of positronium annihilates into two 511~keV gamma photons with a lifetime of 125~ps, during which light propagates less than 4~cm. For the suggested cigar-shaped condensates of 1~cm length, a sufficiently large fraction of the gamma ray photons builds up in the condensate volume and can stimulate emission of annihilation radiation. Here, the stimulation is purely optical, in contrast to the superradiance driven by atomic coherence.

Initial proposals assumed a unitarity limited one-photon stimulation cross section of $\lambda_c^2/2 \pi \approx 10^{-20}$ cm$^2$~\cite{LIANG1988419,MILLS2002,Subharmonic_Mills_2021}. For densities of $10^{20}$~cm$^{-3}$, four orders of magnitude larger than what has been achieved to date~\cite{Cassidy2018Positronium}, the gain would be 1/cm. The maximum density for positronium is possibly limited to 10$^{19}$~cm$^{-3}$ by collisional transitions between the positronium spin states. Beyond this critical density spinor dynamics causes an instability for para-positronium and may limit the gain~\cite{Clark_PhysRevA.89.043624}. Given the limits on density, it was regarded as a major discovery~\cite{Synopsis2014,Mills_Bubbles_PhysRevA.100.063615,Subharmonic_Mills_2021} that the inverse gain length is not proportional to density, but only to the square root, which would provide higher gain at densities below 10$^{20}$~cm$^{-3}$~\cite{Avetissian2014,Avetissian2015,Subharmonic_Mills_2021,Mills_Bubbles_PhysRevA.100.063615}. This result was obtained by correctly modeling the gain as a coherent two-photon stimulation process. However, the authors emphasize the coherence of the BEC~\cite{Avetissian2014}: ``Here we deal with the double coherence involved in two-photon laser emission from a BEC, i.e., coupling of two macroscopic coherent ensembles of bosons—the BEC of Ps atoms and photons''. The derivations~\cite{Avetissian2015, Subharmonic_Mills_2021} imply that this collective and coherent amplification of photon pairs requires the existence of a macroscopic wavefunction and would not occur for a hypothetical ultracold classical (Boltzmann) ensemble of positronium atoms. We disagree. Subharmonic conversion does not necessitate any special properties of a BEC, and it is a coherent process as long as there is no broadening of the gain (see End Matter). For positronium, Doppler broadening of the $\approx$ 1 GHz natural linewidth would occur for velocities of 3 mm/s, corresponding to temperatures of pK --- that's why a BEC is needed (actually a very cold one, since a coherence length of 10 cm~\cite{mills2004} requires temperatures much lower than the transition temperature in order to freeze out phase fluctuations~\cite{Phase_fluct_PhysRevLett.87.160406}).

The higher gain at lower densities does not evade the $\mathrm{OD}=n \lambda^2 l <1$ threshold for stimulated annihilation. The requirement $\mathrm{OD}>1$ was initially derived by assuming a one-photon gain cross section of $\approx \lambda^2$~\cite{LIANG1988419,MILLS2002,Subharmonic_Mills_2021}. In this work, we derived the same threshold only requiring that end fire modes are populated with more than one photon before the whole sample has decayed, which is independent of the gain mechanism. Consequently, enhanced two-photon emission can only develop before the positronium source is depleted if $n \lambda^2 l >1$.
Since the lifetime of Ps limits the useful length of the condensate to a few cm, this requires densities around $10^{20}$~cm$^{-3}$ as already discussed in Ref.~\cite{MILLS2002}. A recent review paper~\cite{Cassidy2018Positronium} concludes that probably even higher densities would be required for stimulated annihilation which can therefore most likely not be observed with existing technologies.

\textbf{Conclusions.}
In conclusion, we have laid out the principles of how Bose-Einstein condensates of radioactive atoms can enhance nuclear decay. 
The advantage of BECs is their extremely small Doppler width given by the momentum uncertainty $\hbar/l$~\cite{Rivlin_2009}. In spite of this, for MeV gamma ray energies, the inverse Doppler width $(l/\hbar)(m/k)$ of the emitted radiation (and therefore the maximum coherence time) is sub-$\mu$s. In addition, the small pm wavelength of MeV radiation requires extremely dense samples to mitigate the simultaneous population of many modes which can deplete the gain medium before stimulation occurs. These two effects lead to a scaling of the gain $g$ with the inverse cube of the gamma ray energy making it virtually impossible to extend the energy range past keV. \footnote{Note that recoil effects can be suppressed in a crystalline solid-state system with recoilless M{\"o}ssbauer lines, where the emitters are tightly localized which is the limit opposite to the BEC case. Due to the Debye-Waller factor, the recoilless line only has reasonable strength for emissions up to 100 keV. It was used in the demonstration of X-ray Dicke superradiance.\cite{chumakov2018_xfel_superradiance}  Here, the multimode competition was avoided by exciting low-lying Dicke state directly with a coherent X-ray laser (and not by spontaneous emission from a fully inverted system, as discussed here) and operating at a Bragg angle.}

\textit{Acknowledgments}.---The authors are grateful to many people for helpful discussions, including Joseph Formaggio, Benjamin Jones, James Thompson, Ana-Maria Rey, Hannes Pichler, David Snoke, Richard Milner, Hamlet Avetissian, Andrei Derevianko and our whole research group. We thank William R. Milner for proofreading the manuscript. Our research has been supported by NSF (grant No. PHY-2208004), from the Center for Ultracold Atoms (an NSF Physics Frontiers Center, grant No. PHY-2317134), by the Vannevar-Bush Faculty Fellowship (grant no. N00014-23-1-2873), by the Gordon and Betty Moore Foundation GBMF ID \# 12405), and by the Army Research Office (contract No. W911NF2410218). This work was performed in part at the Aspen Center for Physics, which is supported by NSF grant PHY-2210452.
 \\

\bibliographystyle{apsrev4-2}
\bibliography{mainbib}{}

\newpage

\section{End Matter}

\subsection{Particle number decay by single-body loss}

Here we show that single-particle loss cannot be modified by any form of many-body physics within a system of $N$ atoms. Using the Lindblad master equation for an arbitrary density matrix \(\rho\):

\[ \frac{d\rho}{dt} = -i[H, \rho] + \sum_i \left( L_i \rho L_i^\dagger - \frac{1}{2} \{ L_i^\dagger L_i, \rho \} \right), \]

where \(L_i = \sqrt{\Gamma} a_i\) are the quantum jump operators for the loss of a single particle in state $i$. This form of loss assumes both the Born and Markov approximations for the reservoir. Note that the decay rate $\Gamma$ can depend on  properties of the environment through the modification of the decay matrix element or the density of states. Examples include the modification of spontaneous decay by the refractive index
and of electron-capture processes by the electron density at the nucleus (see main text). In these cases, the single-particle loss is renormalized by the environment.

The total atom number operator is
\( \hat{N} = \sum_i a_i^\dagger a_i. \)
the evolution of the expectation value of \(\hat{N}\) is expressed as

\[ \frac{d\langle \hat{N} \rangle}{dt} = \text{Tr} \left( \frac{d\rho}{dt} \hat{N} \right). \]

For particle-number-conserving Hamiltonians $H$, 
the term in the Lindblad equation involving $H$ is zero, and we obtain

\[ \frac{d\langle \hat{N} \rangle}{dt} = \Gamma \sum_i \text{Tr} \left( a_i \rho a_i^\dagger \hat{N} - \frac{1}{2} \{ a_i^\dagger a_i, \rho \} \hat{N} \right). \]

Using the cyclic property of the trace, and the commutator of the bosonic operators we obtain
\[ \mathrm{Tr} \left[ \sum_i \left( a_i \rho a_i^\dagger N - \frac{1}{2} (a_i^\dagger a_i \rho N + \rho a_i^\dagger a_i N) \right) \right] \]
\[
= \mathrm{Tr} \left[ \sum_i \rho (a_i^\dagger N a_i - \frac{1}{2} N a_i^\dagger a_i - \frac{1}{2} a_i^\dagger a_i N) \right].  \]

Since
\(  [a_i^\dagger, N] = [a_i^\dagger, a_i^\dagger a_i] = -a_i^\dagger, \quad [a_i^\dagger a_i, N] = 0 \), we find

\[
     \quad \mathrm{Tr} \left[ \rho \sum_i (N a_i^\dagger a_i - a_i^\dagger a_i - \frac{1}{2} N a_i^\dagger a_i - \frac{1}{2} N a_i^\dagger a_i) \right]   \]
\[ = -\mathrm{Tr} (\rho \sum_i a_i^\dagger a_i) 
= -\langle N \rangle  \]
and  \( \quad \frac{d \langle N \rangle}{dt} = -\Gamma \langle N \rangle\) for the atom number decay with the solution
\[ \langle \hat{N}(t) \rangle = \langle \hat{N}(0) \rangle e^{-\Gamma t}. \]
This proves that the total atom number always decays with the single-atom decay rate \(\Gamma\). Note that the proof applies also to very general Hamiltonians $H$ which can include other atomic species B which interact with atom A or even form AB molecules.

Our proof reveals that one has to be careful how to treat loss with jump operators. If we have a system of AB molecules, atom A  decays and all decay products are lost, and one describes this by a jump operator \(\sqrt{\Gamma} a_A\) (loss of A), then our theorem states correctly that there is no change of decay rate due to ``bystander'' atoms like B. This formally explains why the rate of high energy reactions which large recoil cannot be modified by any environment since it acts as a bystander.

This bystander situation could be ideally realized  for  a weakly bound AB molecule close to the dissociation limit, or even more ideally, when the radioactive atom is a para-dimuonium atom which annihilatets into two gamma photons. In this case, the description of decay by annihilation operators for the two muons is exact.

If one would incorrectly describe the decay of atom A (e.g.the dimuonium atom) by a molecular annihilation operator \(\sqrt{\Gamma} a_{AB}\) times a creation operator \(a^\dagger_B\) for B, then the presence of a large condensate of $N$ atoms of type B would lead to bosonic stimulation term of the process $\propto N+1$. The prefactor includes the overlap of the atom B in the localized dissociated state with the condensate which is at best $1/N$, so the stimulation would be small and not scale with $N$. However, our general proof shows that bosonic stimulation is exactly zero since atom B is a bystander. The incorrect description as the loss of a molecule and creation of an atom B implies that atom B has a part in the dynamics of the decay process, since it is created by the ``environment''. In reality, the sudden radioactive decay leads to a reduced density matrix where the lost atom A is traced out without any other modification in the system.

\subsection{Bose-Einstein condensation and emission of light}
Emission of light (or neutrinos) and light scattering do not depend on the coherence of the condensate, but only on the density and Doppler width. These are the only quantities (besides single-particle effects like matrix elements) which affect the coupling of a gas of atoms to light. In a thought experiment, we can replace atoms in a condensate by atoms with the same radiating properties, but make them much heavier. The condensate atoms have momentum spread $\hbar/l$ and velocity spread $\hbar/ m l$. If we use atoms with much higher mass $m^*$ and choose a temperature so the thermal velocity equals $\hbar/ m l$, then their de Broglie wavelength is $l m/m^*$. For sufficiently large $m^*$, the thermal de Broglie wavelength is smaller than the interatomic distance $n^{-1/3}$, and the gas is not condensed. For momentum transfer $k$, by construction, the Doppler width is the same as before $k \Delta v$ = $\hbar k/ml$. Therefore, this thermal ensemble of heavier atoms has the same coherence time and gain as the condensate, demonstrating that any special properties of the condensate do not matter. The Doppler shift is completely suppressed for infinitely heavy atoms, and also by localizing the atoms (e.g. by implanting them into a solid) as in  Mössbauer spectroscopy.

This argument needs to be modified for annihilation radiation where momentum conservation implies that the momentum spread of the photons is identical to the momentum spread of the atoms (divided by two for the emission of photon pairs). However, this is in the end again the Doppler shift: If positronium atoms have a velocity spread $\Delta v$, the annihilation radiation is Doppler broadened by $\Delta \omega = k \Delta v$. The momentum spread of the 511 keV radiation is $\Delta p_\gamma =\Delta \omega \hbar/c$= $m_{e} \Delta v= \Delta p_\textrm{pos}/2$.

Since for annihilation radiation, the momentum spread matters (and not the velocity spread, as for atoms emitting photons), the light will have the same phase-space density as the annihilated atoms, and that makes a BEC unique as a source. However, this does not involve any special coherence properties of the source. In the derivation of the down conversion Hamiltonian describing two-photon annihilation~\cite{Avetissian2014,Avetissian2015,Subharmonic_Mills_2021} one can replace the initial coherent state of positronium atoms by a number state or even by a Boltzmann ensemble of many distinguishable positronium atoms in the ground state of the confinement potential (e.g. by assigning to each atom a different pseudo-color quantum number) and obtain the same coherent gain as long as the source is undepleted. What matters are only the density and momentum distribution --- there is no extra condensate magic!

Although not relevant for the current paper, we want to mention the special case of 2D polaritons or excitons which can annihilate into single photons emitted perpendicular to the 2D plane~\cite{Snoke_doi:10.1126/sciadv.adk6960}. The 2D confinement provides the longitudinal momentum, and the transverse momentum of the light is the momentum spread of the condensate: the photons directly inherit the (transverse) momentum and kinetic energy of the particles which is analogous to the Doppler effect. The brightness of the emitted beam is given by the phase-space density of the excitons. This are the same principles as discussed for positronium annihilation. The difference is that in a single-photon annihilation, the photon field also inherits the phase of the excition wavefunction: When a coherent state of excitons annihilates, then the initial and final states for the single-photon emission are components of the coherent state wavefunction with exciton number $N$ and $N-1$ which, for a condensate in a coherent state $\ket{\alpha}$, have a relative phase identical to the coherent state. Therefore, in this case, the 2D wavefunction of the condensate (including the phase) is mapped to the annihilation radiation. This is different from the positronium two-photon annihilation where entangled photon pairs are created and therefore a single photon does not have a well-defined phase.

This difference could be in principle observed by using metastable condensates as a gain medium for a laser beam: For the examples of BECs of gamma emitters and of positronium, there is gain for an input laser of arbitrary phase, whereas an exciton BEC would provide maximum gain only when the phase of the laser and the phase of the BEC are matched.

We can discuss in a simple Dicke model in the long-wavelength approximation whether a BEC makes a difference. The Hamiltonian is $ {\cal H} = g(ab^{\dagger}c^{\dagger} + \textrm{h.c.})$ 
 where $a$ is annihilation operator of a condensate atom, and $b^{\dagger}$ and $c^{\dagger}$ are creation operators for the recoiling daughter atom and the emitted photon, respectively. For positronium, $b^{\dagger}$ and $c^{\dagger}$ are the operators for the two gamma photons, for exciton annihilation, there is no $b$ particle. This Hamiltonian maps to a spin Hamiltonian for spin $S=N/2$: ${\cal H} = g(S^-c^{\dagger} + \textrm{h.c.})$ for atoms and excitons, and ${\cal H} = g(S^-b^{\dagger}c^{\dagger} + \textrm{h.c.})$ for positronium. The energy levels for the atomic system are equidistant.

If we had identical atoms in the same momentum state as the condensate atoms, but with an additional quantum number $i$, then the Hamiltonian becomes
 \begin{equation}
  {\cal H} = \sum_{i} g(a_i b_i^{\dagger}c_i^{\dagger} + \textrm{h.c.})
\end{equation}

As long as all the atoms emit into the same mode, we can use the collective spin operator
 $S^-= \sum_{i} a_ib_i^{\dagger}$ for atoms, $S^-= \sum_{i} a_i$ for excitons and positronium annihilation and obtain the same Hamiltonians as for the BEC case. This illustrates that any special property of the BEC (macroscopic wavefunction, coherence) is not relevant.

 If we now assume $N$ atoms in the same internal state, but in different motional states $i$, we can map to the same Hamiltonian, as long as the energy differences for the different atomic states $i$ are smaller than the inverse duration of the superradiant pulse $\approx 1/N\Gamma$. In a gas of atoms, these are Doppler shifts $q v_i$. For annihilation radiation, the energy shift of the emitted light is the kinetic energy of the atom (or half of it for positronium). The inverse of these energy shifts define the coherence time. So in the end, there is no qualitative difference between condensed and non-condensed atoms, the quantitative difference is captured by the coherence time of the system.

\makeatother
\onecolumngrid

\end{document}